\documentclass[12pt,preprint,a4paper]{aastex}

\usepackage{graphics,epsf}
\usepackage{amsmath}                
\usepackage{amsfonts}               
\usepackage{amssymb}                
\usepackage{epsfig}                 

\def \eV{~\rm{eV}}

\def \cm{~\rm{cm}}
\def \s{~\rm{s}}
\def \km{~\rm{km}}

\def \K{~\rm{K}}
\def \g{~\rm{g}}

\def \AU{~\rm{AU}}
\def \erg{~\rm{erg}}

\def \yr{~\rm{yr}}

\begin{document}

\title{Evaporation of Jupiter like planets orbiting extreme horizontal branch stars}

\author{Ealeal Bear$^1$ and Noam Soker$^1$}

\affil{1. Department of Physics, Technion -- Israel Institute of Technology, Haifa
32000, Israel; ealealbh@gmail.com; soker@physics.technion.ac.il}

\begin{abstract}

We study the evaporation of planets orbiting close to hot (extreme) horizontal branch (EHB) stars. These
 planets survived the common envelope phase inside the envelope of the reg giant star progenitor. We find
  that Jupiter-like planets orbiting within $\sim 10 R_\odot$ from an EHB star suffers a non-negligible mass-loss
  during their $\sim 10^8 \yr$ evolution on the horizontal branch. The evaporated gas is ionized
     and becomes a source of Balmer lines. Such planets might be detected by the periodic
     variation of the Doppler shift of the Balmer lines.

\end{abstract}

\section{INTODUCTION}
\label{sec:intro}
Horizontal branch (HB) stars are core Helium burning stars that have evolved from
main sequence (MS) stars through the red giant branch (RGB).
During the RGB phase the star loses a non-negligible amount of mass.
The amount of mass lost determines the properties of the descendant HB star.
Namely, its location on the HR diagram.

HB stars with low mass envelope have small radii and they are hot.
They are called extreme HB (EHB) stars (other names are sdO or sdB
or hot subdwarfs; in this work we will use all these terms
indistinguishably). To become an EHB star, the RGB progenitor must
lose most of its envelope. The reason that some RGB stars lose
so much mass was a major unsolved issue in stellar evolution.
The debate was whether a single star (e.g., Yi 2008) can account for
the formation of hot subdwarfs, or whether binary evolution is
behind the hot subdwarf phenomenon (e.g., Han et al. 2007).
Recent studies suggest the binary interaction is behind the formation of
most EHB stars (for a recent paper and more references see Geier et al. 2010a).
However, not all EHB stars have stellar companions.
It has been suggested that massive planet companions can also
influence RGB stars and cause the formation of EHB stars
(Soker 1998; by planet we will refer in this paper also to brown dwarfs).
This model was confirmed with the discovery of a substellar object in a close orbit
to an sdB star (HD 149382, Geier et al. 2009), as well as a planet orbiting a red HB star that lost some of its envelope
(Setiawan et al. 2010).
The intense UV radiation from the EHB evaporates the outer layers of a surviving close in planet.
In this paper we study this process.

The escape of atoms from a planet has been deduced observationally
from absorption of atomic hydrogen around the planet HD209458b
that orbits a MS star (Vidal-Madjar et al. 2003; Vidal-Madjar \& Lecaveleier des Etangs 2004).
In early studies several groups (e.g., Lammer et al. 2003; Baraffe et al. 2004, 2005)
suggested that hot Jupiters orbiting MS stars can be evaporated down to their bare core.

Many detailed calculations have been made on evaporation of
planets in different conditions and circumstances (Dopita \&
Liebert 1989; Schneider et al. 1998; Schneiter et al. 2007; Soker
1999; Lammer et al. 2003; Baraffe et al. 2006; Erkaev et al. 2007;
Jackson et al. 2008; Garcia Munoz 2007; Lammer et al. 2009;
Murray-Clay et al. 2009). Villaver \& Livio (2007), for example,
calculated the outflowing particle flux by equating the energy
input and the energy required for hydrogen to escape. Their
treatment is not much different from those of others (e.g.,
Baraffe et al. 2004; Erkaev et al. 2007; Lecavelier des Etangs
2007,Lecavelier des Etangs et al. 2008; Penz et al. 2008a; Lammer et al. 2003, 2009; Valencia
et al. 2010a; Sanz-Forcada et al. 2010). Another approach which takes
into account the recombination of the evaporated gas is presented
by Dopita \& Liebert (1989) and McCray \& Lin (1994).

Different models predict different mass-loss rates (e.g., Hunten
1982; Sasselov 2003; Vidal-Madjar \& Lecavelier des Etangs 2004;
Erkaev et al. 2007; Hubbard et al. 2007; Ehrenreich 2008,
Ehrenreich et al. 2008; Davis \& Wheatley 2009; Lammer et al. 2009
\& Linsky et al. 2010).
Murray-Clay et al. (2009) comprehensively review the basic ``energy - limited''
model that is based on channelling heating radiation to mass-loss.
In the simplest approach most of the ionizing radiation energy goes into work to expel the envelope.
This model is similar to the one used by Lecavelier des Etangs (2007), but the assumption of $100 \%$ conversion
is unrealistic and overestimate the mass-loss rate.
A more realistic approach limits the radiation energy available for mass-loss.
In their model Murray-Clay et al. (2009) take a realistic heating efficiency of $10 - 30\%$, since not
all the absorbed EUV energy is channelled into heating.
Other hydrodynamical models by Yelle (2004), Garcia Munoz (2007),
 Erkaev et al. (2007) and Lammer et al. (2009)
take the same approach. Soker (1999; based on Dopita \& Leibert 1989), for example,
further took into account the recombination of the outflowing gas.
This process causes a decrease in the mass-loss rate.
We will use the energy-limited process with $10-30 \%$ efficiency.
For example, by considering the effect of recombination of the outflowing gas.
This makes the model generally applicable to high and low ionization
fluxes for planets around EHB stars.

Lecavelier des Etangs et al. (2004) and Lecavelier des Etangs (2007) concluded based on their detailed
calculations that planets with orbital distances of $0.03-0.04
\AU$ from a MS star will be evaporated unless they are
significantly heavier than Jupiter. This approach is strengthened
by Davis \& Wheatley (2009) who examine the EUV from MS stars (F,
G and K), and conclude that planets will not exist at small orbital distances. Let us mention
a number of observed cases of planets orbiting MS stars, that
motivate our study of planets orbiting HB stars, in particular EHB
stars. Valencia et al. (2010a, b) raised the possibility that the
super-earth like planet CoRot-7~b ($M_p=4.8 \pm 0.8M_\oplus$,
$R_p=1.68 \pm 0.09R_\oplus$, $a_p=0.017\AU$, $e\sim 0$) is the
outcome of evaporation of an Uranus like planet. Baraffe et al.
(2004) find that a planet with a mass below a critical mass of
$m_{\rm crit}=2.7M_{\rm J}$ orbiting a solar-type star at an
orbital separation of $a_p=0.023\AU$, will be completely evaporated
in $5~$Gyr, unless it has a central rocky core. Jackson et al.
(2010) elaborated on the importance of evaporation and calculated
two paths. In the first CoRoT-7~b has always been a rocky planet,
and in the second CoRoT-7~b is a remnant of a gas giant. Jackson
et al. (2010) took into consideration tides, and concluded that it
is possible that CoRoT-7~b is a remnant of a gas giant planet. If
this finding holds to the cases we study here, it is possible that future observations will
reveal many more ``earth like planets'' around white dwarfs (WDs) or HB stars,
that actually started their life as gas giant planets.

We start by studying the evaporation of planets orbiting EHB stars
(sec. \ref{sec:loss}). The gas escaping from the planet will be
ionized by the radiation of the HB parent star, and become a
source of H$\alpha$ emission. This idea has been raised before as
an indirect way to search for planets in Planetary Nebulae and
Jupiter like planets around WDs (Soker 1999; Chu et al. 2001). We
modify this idea and try to search for planets around EHB stars
through their H$\alpha$ emission. In section \ref{sec:halpha} we
examine the conditions for this emission to be detected. Our short
summary is in section \ref{sec:summary}.

\section{EVAPORATION OF A PLANET ORBITING AN HB STAR}
\label{sec:loss}
\subsection{Basic evaporation processes}
\label{sec:loss1}

We start by considering heating by EUV radiation, a process
that was studied in detail for MS and pre-MS central stars (e.g.,
Chamberlain \& Hunten 1987; Yelle 2004; Tian et al. 2005). At this
stage we will not consider the role of the magnetic field of the
planet, although it can play some role (e.g., Griebmeier et al.
2004; Lammer et al. 2009). We adopt the simple model presented by
Lecavelier des Etangs (2007) which represents the blow - off mechanism (Erkaev et al. 2007)
 and investigate the implications for
a planet orbiting an HB star (this model is similar the model purposed by Murray-Clay et al. 2009). The potential
energy per unit mass in the atmosphere is
\begin{equation}
\frac{dE_{p(\rm atm)}}{dm}= \frac{GM_p}{R_p}= \frac{v_{\rm esc}^2}{2}= -1.8\times 10^{13}
\left( \frac{M_p}{M_{\rm J}} \right)
\left( \frac{R_p}{R_{\rm J}} \right)^{-1}   {\rm erg}~ g^{-1},
\label{eq:ep}
\end{equation}
where $M_p$, $R_p$, $M_{\rm J}$ and $R_{\rm J}$ are the planet
mass, planet radius, Jupiter mass, and Jupiter radius
respectively and $v_{\rm esc}$ is the escape velocity from the planet.
Even for very-hot Jupiters the magnitude of the potential energy is much
larger than the kinetic energy of thermal gas particles, and we follow Lecavelier des
Etangs (2007) and neglect the kinetic energy of atoms in the planet atmosphere.

The general expression for mass-loss according to Lecavelier des Etangs (2007), is
\begin{equation}
\dot m_p =\frac{2\eta \dot E_{\rm EUV}}{v_{\rm esc}^2},
\label{Eq.2}
\end{equation}
where $\dot E_{EUV}$ is total EUV power in the range of
$100\AA \leq \lambda\leq 1200\AA$ (Lecavelier des Etangs 2007) received by the planet.
We took into account that not all the absorbed EUV radiation will be channelled to evaporation
by introducing the parameter $\eta \simeq 0.1-0.3$.
Although some studies use $\eta =1$ (e.g. Lammer et al. 2003; Baraffe et al. 2004;
Lecavelier des Etangs 2007), more recent studies found the efficiency to be lower,
e.g., Penz et al. (2008b) find $\eta <0.6$ for hydrogen rich thermosphere,
and Lammer et al. (2009) find $\eta \simeq 0.1-0.25$.
Most significant in reducing the efficiency is L$\alpha$ cooling by collisionally excited hydrogen atoms
(Murray-Clay et al. 2009).

\begin{figure}
\includegraphics[scale=0.8]{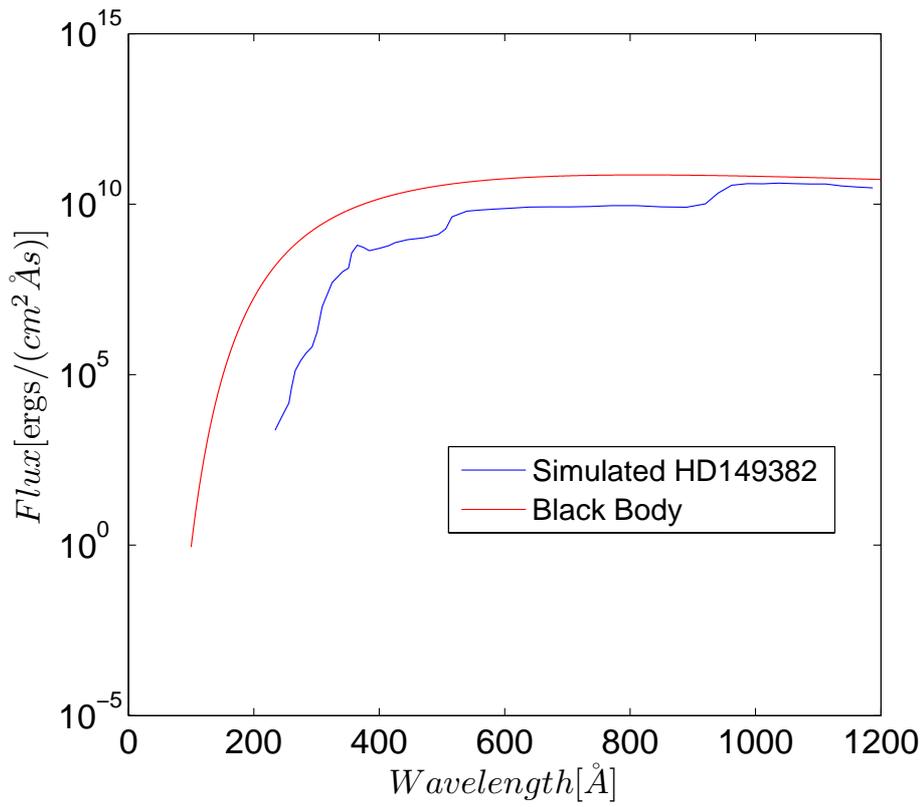}
\caption{The spectrum $[{\rm erg }\cm^{-2} \s^{-1} \AA^{-1}]$. The red (upper) line represents
the flux of the black body. The blue line represents the flux of the
simulated HD149382.}
\label{fig:Compare}
 \end{figure}

An appropriate calculated spectrum is required for EHB stars since a black body (BB) radiation does not
fit the spectrum below $912 \AA$.
In Figure \ref{fig:Compare} we compare the spectrum calculated by Geier et al. (2010b) for HD 149382,
an sdB star with an effective temperature of $T=35,500 \K$ and $\log(g)=5.75$, where $g~(\cm \s^{-2})$ is
the gravity on the stellar surface, with a BB radiation at the same temperature.
In the case of a BB radiation we have
\begin{equation}
\dot E_{\rm EUV}
=\pi R_p^2 \frac{R_{\rm EHB}^2}{{a_p}^2}
\int_{100\AA}^{1200\AA}
\frac{2\pi h c^2/\lambda^5}{\exp(hc/\lambda k T)-1}d \lambda,
\label{Eq.3}
\end{equation}
where $h$ is the planck constant, $c$ is the speed of light, and $k$ is the Boltzmann constant.

Soker (1999, where more details are given) calculates the mass ablation rate of the planet
by taking the ionization approach,
but including the effect of recombination, following McCray \& Lin (1994) who calculated
the ablation of the ring around SN1987A.
Recombination transfers kinetic energy to radiation that
escapes, and reduces the ablation rate.
The ionization rate is multiply by the ratio of recombination time to
escape time (as long as this ratio is not larger than 1).
The expression derived by Soker (1999) is
\begin{equation}
\dot m_p  \simeq N\eta_i \mu m_H \left( \frac{\tau}{n} \right)
\left( \frac{R_p}{c_s} \right)^{-1},
\label{Eq.dmdt1}
\end{equation}
where $\tau/n$ is the recombination time, $n$ is the total number density of the ablated
layer, $R_p/c_s$  is the escape time from the planet, $c_s$ is the speed of sound, $N$ is the rate of ionizing photons
hitting the planet, $\eta_i \simeq 0.1$ is the ionization efficiency and $\mu m_H$ is the mean mass per particle.
The ionizing rate is given by $N=N_* {\left(\frac{R_p}{2a_p}\right)}^2$, where $N_*$
is the number of ionizing photons per unit time emitted by the HB star (Soker 1999).
Assuming that the evaporated mass outflows at the sound speed and toward the half hemisphere facing the star, the mass-loss rate is
\begin{equation}
\dot m_p \simeq 2\pi n \mu m_H R_p^2 c_s.
\label{Eq.dmdt2}
\end{equation}
We eliminate $n$ from equations (\ref{Eq.dmdt1}) and (\ref{Eq.dmdt2}) and obtain
\begin{equation}
\dot m_p \simeq  2\pi c_s \mu m_H R_p^{1.5} a_p^{-1}\sqrt{\frac{\tau N_* \eta_i}{8 \pi }}.
\label{Eq.dmdt3}
\end{equation}

It must be emphasized that the ionization evaporation rate given by Eq.
(\ref{Eq.dmdt3}) was used by Soker (1999) for Uranus like planets,
that have very low escape energy (Eq. \ref{eq:ep}).
For more massive planets the escape energy is comparable to the energy of the ionizing
radiation, and cannot be neglected. Therefore, the evaporation rate given by Eq.
(\ref{Eq.dmdt3}) becomes inappropriate when it gives value above that given by
Eq. (\ref{Eq.2}).
In this paper we deal with massive planets and with brown dwarf orbiting close to HB stars.
We consider the ionization evaporation rate as a cautionary step, because it
takes into account recombination that reduces the efficiency.

\begin{figure}
\includegraphics[scale=0.8]{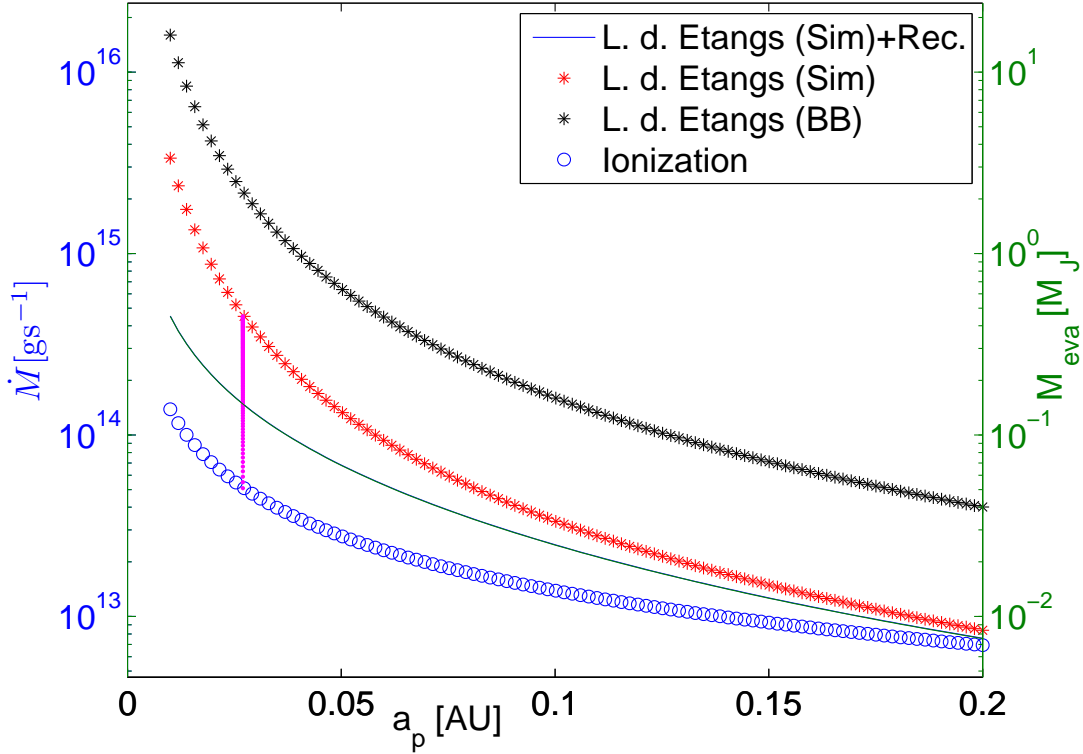}
\caption{Mass evaporation rate $\dot m_p$ (left axis) versus the orbital separation $a_p$.
The right axis gives the total mass that would be evaporate during a period of $6 \times 10^7 \yr$.
The calculated mass-loss curves were done for $\eta =\eta_i= 0.1$. The blue circles (lower line) represent the ionization model from equation (\ref{Eq.dmdt3}).
The black thick (upper) line represents the evaporation rate based on  Lecavelier des Etangs (2007)
as given by equation (\ref{Eq.2}) for a black body energy distribution (\ref{Eq.3}).
The red thick line represents the evaporation rate based on Lecavelier des Etangs (2007) for a correct spectrum of
HD 149382 (Geier et al. 2010b).
The blue thin line represents the same model of Lecavelier des Etangs (2007) using the correct spectrum, but with
recombination of the evaporated gas included (equation \ref{eq:dotmp}). This last case (blue thin line) is the appropriate case to use,
and the one used in calculating the equivalence width of the H$\alpha$ and H$\beta$ emission lines in section \ref{sec:halpha}.
The evaporation rates are calculated for an EHB central star and a planet with the properties
of the HD 149382 system:
$T_{\rm EHB}=35500K$, $M_{\rm EHB}=0.5M_\odot$, $R_{\rm EHB}=0.14R_\odot$, $M_p=15M_{\rm J}$,
(Geier et al. 2009), and  $R_p=0.1R_\odot$.
The orbital separation of this system is $a_p=5-6.1R_\odot$, but here it is an independent variable.
The magenta line represents an orbital separation of $a_p=5.8 R_\odot$. }
\label{fig:ML}
 \end{figure}

Fig. \ref{fig:ML} presents the ablation rate based on Lecvelier des Etangs (2007) as given by equation
(\ref{Eq.2}), with the ionization model (Dopita \& Liebert 1989; Soker 1999)
as given here by equation (\ref{Eq.dmdt3}), both as
function of the orbital separation.
These are calculated with the appropriate spectrum as was calculated for HD 149382 (Fig. \ref{fig:Compare}).
For comparison we show the evaporation rate for a BB spectrum with the same
effective temperature and luminosity (black upper line).
The ionization model is presented in figure Fig. \ref{fig:ML} only for comparison purposes and it does
not apply when the escape velocity exceeds the sound speed.

The properties of the EHB
central star and the planet are taken to be those of the HD 149382
system (Geier et al. 2009; see figure caption). The orbital
separation of this system is $a_p=5-6.1R_\odot$, but in the figure
this is an independent variable. On the right axis of Fig.
\ref{fig:ML} we give the total mass that would be evaporate during
a period of $6\times 10^7 \yr$, about the duration of the HB,
with the same mass-loss rate given on the left axis. For these
parameters we find $\dot E_{\rm EUV} = 4.4\times 10^{30} \erg
\s^{-1}$ and $N \sim 3.8\times 10^{40} \s^{-1}$ and we assume an
 efficiency of $\eta=0.1$, and $v_0 \simeq c_s \simeq 10 \km \s^{-1}$, where $v_{\rm o}$ is the outflow velocity.
 For the ablation rate based on ionization (equation \ref{Eq.dmdt3}) we substitute the
following numerical values: $\tau=3\times 10^{12} \s$ (Osterbrock
1989) and $\mu=0.62$. The expanding gas does not reach the escape velocity.
It escapes the planet when it leaves the planet's Roche lobe. The mass-loss curves in Fig. \ref{fig:ML} were
calculated for $\eta=\eta_i = 0.1$, and therefore represent a lower limit.
For the evaporation process to be efficient, the orbital separation cannot be too large,
i.e., $a \la 0.1 \AU$, depending on the exact planet properties (Davis \& Wheatley 2009 and references therein).
We here show the results up to an orbital separation of $0.2\AU$.

 We now turn to include recombination in the energy-limited
process, as this is the more realistic approach. We do it for the
parameters of HD 149382 (represented by the blue thin line in Fig.
\ref{fig:ML}).

\subsection{Including recombination of the evaporated gas}
\label{sec:recombin}

When the central source is hot a large fraction of the radiation is energetic enough to
ionize the evaporated gas. The evaporated gas recombines and emits at a longer wavelength radiation that
escapes from the planet's vicinity. Although recombination is not relevant to planets
around solar-like stars, its role becomes more important for hot HB stars and central stars
of planetary nebulae.
To facilitate a simple calculation we make the following simplifying assumptions.
\begin{enumerate}
\item Most of the evaporated gas flows toward the radiation source, i.e., the parent star.
Namely, the evaporated gas escapes to a solid angle of $4 \pi \beta$ with $\beta=0.5$.
\item The central star keeps the gas almost fully ionized, such that the rate of recombination equals
    that of ionization by the radiation of the parent star.
\item The ionizing photons of the parent star that are absorbed by the evaporated gas
    are removed from the radiation that heat the star.
\item Most of the recombination radiation is by gas close to the planet where density is high.
     Therefore, a half or less of the radiation of the recombining evaporated gas
     will be absorbed back by the planet and heat it. To put an upper limit on the role
      of recombination, we assume that all the radiation emitted by the recombining gas
      escapes.
\item We assume that the gas outflow velocity is about equal to the sound
        speed $\sim 10 \km \s^{-1}$ (Gu et al. 2003; Li et al. 2010; Lai et al. 2010;
        Trammell et al. 2010 and references therein).

\end{enumerate}

The recombination rate is proportional to the density square, hence the square of the mass-loss rate.
Therefore, the rate the evaporated gas removes photons from the parent stellar radiation is
$\dot N_{\rm rec} = K_2 \dot m_p^2$, where $K_2$ is a constant to be derived below.
Instead of equation (\ref{Eq.2}), the new equation reads now
\begin{equation}
\dot m_p =\eta \frac{2}{v_{\rm esc}^2}
\left( \dot E_{\rm EUV} - K_2 e_\gamma \dot m_p^2 \right)
=\eta \dot m_{p0} - \eta \frac{ 2K_2 e_\gamma } {v_{\rm esc}^2}  \dot m_p^2,
\label{eq:dotmp}
\end{equation}
where $e_\gamma \sim 20 \eV$ is the average energy of the ionizing photons,
and in the second equality we defined the zeroth order evaporation rate (when recombination is neglected and $\eta=1$)
$\dot m_{p0} = 2\dot E_{\rm EUV} / v_{\rm esc}^2$.
Equation (\ref{eq:dotmp}) is a quadratic equation that can be solved analytically.
By our assumptions, the density of the evaporated gas is
\begin{equation}
\rho = \frac{ \dot m_p } { 4 \pi \beta v_{\rm o} r^2},
\label{eq:density}
\end{equation}
where $v_{\rm o}$ is the outflow velocity which is taken as $10\km \s^{-1}$.
The recombination rate per unit volume is $\dot n_{\rm
rec}=\alpha_{\rm rec} n_e n_p$, where by the assumption of
(almost) fully ionized gas can be written as $\dot n_{\rm
rec}=\bar \alpha_{\rm rec} \rho^2$, where $\bar \alpha_{\rm rec} =
5 \times 10^{34} \cm^3 \g^{-2} \s^{-1}$ is appropriately
calculated from $\alpha_{\rm rec}$ for a fully ionized solar
composition in case B recombination (Osterbrock 1989). We neglect
processes that become more important due to the high collision
rate expected in the very dense outflowing gas near the planet.
The total recombination rate is obtained by integrating over the
entire volume according to our assumptions
\begin{equation}
\dot N_{\rm rec}=\int_{R_p}^{\infty}  \bar \alpha_{\rm rec} \rho^2  4 \pi \beta r^2 dr.
\label{eq:rec1}
\end{equation}
Substituting equation (\ref{eq:density}), and performing the integration gives
\begin{equation}
\dot N_{\rm rec}= K_2 \dot m_p^2 =
\frac {\bar \alpha_{\rm rec}}{ 4 \pi \beta v_{\rm o}^2 R_p} \dot m_p^2.
\label{eq:rec2}
\end{equation}
The last equality gives the value of $K_2$ that we substitute into equation (\ref{eq:dotmp}).

Recombination becomes important when the last term in equation (\ref{eq:dotmp})
becomes non negligible. Taking $\dot m_p \simeq  \dot m_{p0}$, this occurs when
\begin{equation}
\dot m_{p0} \ga \frac{v_{\rm esc}^2} {2K_2 e_\gamma }
= \frac { 2 \pi \beta v_{\rm o}^2 v_{\rm esc}^2 R_p} {\bar \alpha_{\rm rec} e_\gamma}.
\label{eq:dotmp3}
\end{equation}
Substituting typical values gives the evaporation rate above which recombination is important
\begin{equation}
\dot m_{p0} \ga  9\times 10^{12} 
\left( \frac{\beta}{0.5} \right)
\left(\frac {v_{\rm esc}} { 250 \km \s^{-1}} \right)^2
\left( \frac {v_{\rm o}} { 10 \km \s^{-1}} \right)^2
\left( \frac{R_p}{0.1 R_\odot} \right)
\left( \frac{e_\gamma}{20 \eV} \right)^{-1} \g \s^{-1}.
\label{eq:dotmp4}
\end{equation}

In Fig. \ref{fig:ML} the energy-limited process is
included with recombination (equation \ref{eq:dotmp}) and is
depicted by the blue thin line. It can be seen that the
recombination becomes important when the evaporation rate is as
given in equation (\ref{eq:dotmp4}). Namely, it is important in
the entire relevant range of parameters here. The
evaporation rate we will use in calculating the H$\alpha$ emission
is the one given by the blue thin line of Fig. \ref{fig:ML}.

The substellar object (a planet or a BD) mass in HD 149382 is
$8-23 M_{\rm J}$ (Geier et al. 2009) at an uncertain orbital
separation of $a_p =5-6.1 \AU$.
From Fig. \ref{fig:ML} we learn that the total evaporated mass of this object during the HB phase will be
$\sim 0.1-1 M_{\rm J}$. This amount is significant,
but seems that the substellar object in this system will survive the HB phase of its parent star.
\section{H$\alpha$ EMISSION OF THE EVAPORATE MATERIAL}
\label{sec:halpha}

We consider here hot HB stars such that the evaporated gas of close planets
is almost fully ionized. The calculation of the H$\alpha$
luminosity from the evaporated gas is done in the following way
(e.g. Bhatt 1985 for destructed comets). We start with the
following assumptions, some of which were used in section
\ref{sec:loss}.
\begin{enumerate}
\item The evaporation is mainly into a solid angle $4 \pi \beta$.
If it is toward the parent star $\beta \simeq 0.5$, while if it
is spherical $\beta=1$.
\item Close to the planet, where most of the recombination occurs, the material flows
at the sound speed.
\item For typical values we find the medium to be optically thin to H$\alpha$.
\item We assume that the evaporated gas is almost completely
ionized. Any recombination that occurs is balanced by the incoming
photons from the EHB star. \item Most of the recombination and the
H$\alpha$ source occur at a relatively high density of $n \simeq
10^{10}-10^{12} \cm^{-3}$. At such densities collision between
atoms change the amount of energy that is channelled to
H$\alpha$. In our simple treatment we neglect the dependence of the recombination coefficient
on density. We note that
Bhatt (1985) calculates the H$\alpha$ emission from a destructed
comet. He estimates the density to be $\sim 10^{13} \cm^{-3}$ and
neglects the dependence on density. Korista et al. (1997) found
that the dependence of the recombination coefficient to H$\alpha$ on density in these densities is
negligible.
\end{enumerate}

The H$\alpha$ energy released due to recombination is:
\begin{equation}
L_{\rm H\alpha} = \int_{R_p}^{\infty}\alpha_H (h\nu_{H\alpha})n_e n_p dV
\label{Eq.Ha1}
\end{equation}
Solving the integral yields
\begin{equation}
L_{\rm H\alpha} \sim 2 \times 10^{28}   
\left(\frac{\dot M}{10^{14}\g \s^{-1}}\right)^2
\left( \frac{\beta }{0.5} \right)^{-1}
\left(\frac{R_p}{0.1R_\odot}\right)^{-1}
\left(\frac{v_{o}}{10 \km \s^{-1}}\right)^{-2}   \erg \s^{-1} ,
\label{Eq.Ha2}
\end{equation}

The equivalent width of the H$\alpha$ emission
is calculated for the simulated (accurate)
spectrum of HD 149382 (Geier et al. 2010b), where $T_{\rm EHB} = 33500K$, $R_{\rm sdB}=0.14R_\odot$,
$a_p=0.027\AU~ (5.8R_\odot)$.
When assuming heating efficiency of $\eta = 10\%$, $\dot M=1.5\times 10^{14} \g \s^{-1}$ therefore, we get
$L_{{\rm H} \alpha} = 3.6 \times 10^{28} \erg \s^{-1}$ and hence $EW_{\alpha} \sim 0.09 \AA$ for H$\alpha$ emission and
 $EW_{\beta} \sim 0.01 \AA$ for $H\beta$ emission. The expected  H$\alpha$ emission is within the capability of existing telescopes,
 while the expected H$\beta$ emission seems to be below detection limit.
 When changing the heating efficiency to $\eta = 30\%$,the mass-loss becomes $\dot M=1.7\times 10^{14} \g \s^{-1}$
 and we get $EW_{\alpha} \sim 0.1 \AA$ for H$\alpha$ emission and
 $EW_{\beta} \sim 0.014 \AA$ for $H\beta$ emission.
Although the EWs are not high in both cases, their periodic variation might ease the detection of the line.
At an orbital separation of $5.8 R_\odot$ the orbital velocity of the substellar companion is $\sim 130 \km \s^{-1}$.
Therefore, during the orbital period the center of the emission by the evaporated gas might move back and forth
over a range of up to $\sim 5.5 \AA$ and $\sim 4.0 \AA$, for the H$\alpha$ and H$\beta$ emission lines, respectively.
We conclude that it might be possible to identify a planet via the H$\alpha$ emission of its
ablated envelope.

\section{SUMMARY}
\label{sec:summary}

We estimated the evaporation mass-loss rate from a planet heated
by its parent hot sdB/sdO (EHB) star. The hot star ionizes the
evaporated gas. We assume that it is almost fully ionized. We
reconcile two known evaporation mechanisms (summarized in section
\ref{sec:intro}) by including the effect of recombination in the
evaporated gas, and using the energy-limited model. We
then calculated the expected emission in the lines of H$\alpha$
(equation \ref{Eq.Ha2}) and H$\beta$. As the emission comes from
the planet vicinity, the Doppler shift will be of tens of $\km
\s^{-1}$ over the orbital period. The emission  with its periodic
Doppler shift can be used to directly detect the planet. We note
that Bhatt (1985) proposed to observed the H$\alpha$ emission from
destructed extra-solar comets.

We found that for the substellar object of the system HD 149382 (Geier et al. 2009) the
equivalence widths of the emission of the two lines might be as high as
$EW_{\alpha} \simeq 0.1 \AA$ and $EW_{\beta} \simeq 0.01 \AA$, respectively, and the
Doppler shifts will periodically vary on a range of up to $\sim 5.5 \AA$ and $\sim 4.0 \AA$, respectively
(depending on the inclination of the system).
The detection of the lines is not simple (in particular H$\beta$), as the EHB star itself has absorption in
those lines. However, the periodic Doppler variations might help recognize the emission lines by the
evaporated gas from the planet.

The total evaporated mass along the HB evolution can be non-negligible. However, we can assume
(despite the big uncertainties) that the planet in HD 149382 will survive the entire HB evolution of the star.

The ramification of our study is that sdB/sdO (EHB) stars should
be a prime target for high spectral resolution observation in the
H$\alpha$ (equation \ref{Eq.Ha2}) and H$\beta$ lines. The
observation should look for Doppler variations with an amplitude
of tens of $\km \s^{-1}$, with a period of hours to weeks, that
hint to the presence of an evaporating planet. The target stars are
sdB/sdO stars in the field (disk of the galaxy), where metallicity
is higher. EHB stars in globular clusters are less likely to have
surviving sub-stellar objects, and they are typically at large
distances. Still, some fraction of EHB stars in globular clusters
might have surviving substellar objects around them.

We thank Stephan Geier and Uli Heber for helpful discussions and
suggestions. We thank the referee for very helpful comments. The
Research was supported in part by the N. Haar and R. Zinn Research fund at the Technion,
The Israel Science Foundation, and The Center for Absorption in Science, Ministry of Immigrant Absorption,
State of Israel.

\end{document}